\documentclass[12pt]{iopart}
\usepackage{iopams}
\usepackage{setstack}
\usepackage{graphicx,epstopdf}

\usepackage[colorlinks,linktocpage,linkcolor=blue]{hyperref}

\begin{document}

\title[]{%Superdiffusion of weakly damped Langevin equation in unconfined or confined potential
L\'{e}vy-walk-like Langevin dynamics}

\author{Xudong Wang, Yao Chen, and Weihua Deng}

\address{School of Mathematics and Statistics, Gansu Key Laboratory of Applied Mathematics and Complex Systems, Lanzhou University, Lanzhou 730000, P.R. China}
\ead{xdwang14@lzu.edu.cn, ychen2015@lzu.edu.cn, and dengwh@lzu.edu.cn}
\vspace{10pt}
%\begin{indented}
%\item[]August 2017
%\end{indented}

\begin{abstract}
Continuous time random walks and Langevin equations are two classes of stochastic models for describing the dynamics of particles in the natural world. While some of the processes can be conveniently characterized by both of them, more often one model has significant advantages (or has to be used) compared with the other one. In this paper, we consider the weakly damped Langevin system coupled with a new subordinator|$\alpha$-dependent subordinator with $1<\alpha<2$. We pay attention to the diffusion behaviour of the stochastic process described by this coupled Langevin system, and find the super-ballistic diffusion phenomena for the system with an unconfined potential on velocity but sub-ballistic superdiffusion phenomenon with a confined potential, which is like L\'{e}vy walk for long times. One can further note that the two-point distribution of inverse subordinator affects mean square displacement of this coupled weakly damped Langevin system in essential.
\end{abstract}

\section{Introduction}

Introduced just 100 years ago for describing Brownian motion \cite{Langevin:1908}, the Langevin equation, nowadays, has more wide applications and plays a central role in modeling the dynamical systems coupled with a fluctuating environment \cite{CoffeyKalmykovWaldron:2004}. One of the main advantages of this equation is that it builds a relation between physically transparent and mathematically tractable description for a complex stochastic dynamical system. Assuming that a particle moves in a fluid without friction, it receives a blow due to a random collision with a molecule, then the velocity of
the particle changes. This procedure can be well modeled by weakly damped Langevin system. However, if the fluid is very viscous, the change of velocity is quickly dissipated and the net result of an impact is a change in the displacement of the particle. The overdamped Langevin system becomes more suitable to describe the motion of particles in this case.

Another kind of popular microscopic model is continuous time random walk (CTRW), originally introduced by Montroll and Weiss in 1965 \cite{MontrollWeiss:1965}. It is a powerful mathematical framework to model complex dynamical behaviors, especially anomalous diffusion  phenomena characterized by nonlinear time dependence of mean squared displacement (MSD); see the reviews \cite{HausKehr:1987,BouchaudGeorges:1990,MetzlerKlafter:2000} and references therein. In the CTRW framework, the motion of particles is described through consecutive waiting times and jumps between them. The two random variables, waiting times and jump lengths, are drawn from some associated distributions, which could decide the diffusive behaviour of the particles.
Fogedby \cite{Fogedby:1994} proposed that an overdamped Langevin equation in operation time $s$ coupled with a physical time process $t(s)$ (named as a subordinator) can model the same process as CTRWs in scaling limits. There is an advantage in the Langevin system that the external force field can be included  naturally with clear physical meaning, and the system is given as
\begin{equation}\label{model2}
  \frac{{\rm d}}{{\rm d} s}x(s)=f(x)+\xi(s), \quad \frac{{\rm d}}{{\rm d} s}t(s)=\eta(s),
\end{equation}
where the position $x$ is penalized by the operation time $s$. When $f(x)=0$ and $\xi(s)$ is Gaussian white noise, model (\ref{model2}) yields subdiffusion. 
%In this model (\ref{model2}), position $x$ is penalized by the operation time $s$. If $t(s)$ is $\alpha_0$-stable subordinator ($0<\alpha_0<1$), it yields subdiffusion when $f(x)=0$ and $\xi(s)$ is Gaussian white noise. 
Nowadays, subordinator is a very effective tool to characterize various complex dynamical systems, especially some real-life data in biology \cite{GoldingCox:2006}, financial time series \cite{JanczuraOrzelWylomanska:2011}, ecology \cite{Scheretal:2002}, and physics \cite{NezhadhaghighiRajabpourRouhani:2011}. Besides, overdamped Langevin equation together with its generalizations are well-developed in recent years, e.g., the heterogeneous diffusion processes \cite{CherstvyChechkinMetzler:2013,CherstvyMetzler:2013,CherstvyMetzler:2014} and Brownian yet non-Gaussian diffusion \cite{ChechkinSenoMetzlerSokolov:2017,SposiniChechkinSenoPagniniMetzler:2018}.

At the same time, there are also many processes in practice which could not be well characterized by model (\ref{model2}), since the particles may be weakly damped, e.g., cold atoms diffusing in optical lattices \cite{KesslerBarkai:2012,BarkaiAghionKessler:2014}, and the class of viscoelastic diffusion described by the generalized Langevin equation with (tempered) power-law friction kernel \cite{Lutz:2001,Goychuk:2012,SlezakMetzlerMagdziarz:2018,DengBarkai:2009} and of (tempered) fractional Brownian motion \cite{MandelbrotNess:1968,MeerschaertSabzikar:2013,ChenWangDeng:2017}.
Another main class of generalized weakly damped Langevin equations are coupled with a subordinator not a friction kernel. Eule {\it et al.} \cite{EuleFriedrichJenkoKleinhans:2007} presented three kinds of weakly damped Langevin system coupled with the $\alpha_0$-stable subordinator ($0<\alpha_0<1$), which are related to three kinds of fractional Klein-Kramers equations \cite{MetzlerKlafter2:2000,BarkaiSilbey:2000,FriedrichJenkoBauleEule:2006}, respectively.
In this paper, we also consider the weakly damped Langevin system, but extend the subordinator to be $\alpha$-dependent with $1<\alpha<2$. To the best of our knowledge, the subordinator with $1<\alpha<2$ has never been considered in Langevin system before. One possible difficulty is that the original $\alpha$-stable L\'{e}vy process with $1<\alpha<2$ is not a non-decreasing random process while the one-sided  $\alpha_0$-stable with $0<\alpha_0<1$ is. The condition of non-decreasing must be guaranteed from a physical point of view since the subordinator $t(s)$ denotes the waiting time process in CTRWs \cite{CairoliBaule:2017}.
Fortunately, through L\'{e}vy-Khinchin representation \cite{Applebaum:2009}, an appropriate subordinator can be designed by specifying a L\'{e}vy measure. The method of generating this  subordinator for numerical simulations is given in the last part.

Based on the designed $\alpha$-dependent subordinator with $1<\alpha<2$, we mainly discuss the diffusive behaviour of the weakly damped Langevin system coupled with this subordinator and with two different potentials, i.e., unconfined and confined ones on velocity. The harmonic potential $U(v)=\gamma v^2/2$ is chosen with $\gamma=0$ and $\gamma\neq0$, respectively, corresponding to the unconfined and confined case.
% to represent these two cases with $\gamma=0$ and $\gamma\neq0$, respectively. 
 For long times, in the former case, the particles spread like Richardson-Obukhov diffusion in turbulence, where the velocity follows a simple Brownian motion \cite{Obukhov:1959,BauleFriedrich:2006}. In the latter case, the particle motion is like L\'{e}vy walk \cite{ZaburdaevDenisovKlafter:2015,SanchoLacastaLindenbergSokolovRomero:2004,RebenshtokDenisovHanggiBarkai:2014,ZaburdaevDenisovHanggi:2013} in the sub-ballistic superdiffusion regime. In this way, the mathematical description of L\'{e}vy walk confined to an external force field could be constructed naturally. It is discovered that the essential difference made by the new subordinator on the diffusive behavior
 % We point out the essential difference the new subordinator makes on diffusive behaviour 
 comes from the two-point probability density function (PDF) of inverse subordinator.

The remainder of this paper is organized as follows. In section \ref{Sec2}, we define the $\alpha$-dependent subordinator ($1<\alpha<2$), and discuss its properties as well as its corresponding inverse subordinator. In section \ref{Sec3}, we present the weakly damped Langevin system coupled with this subordinator, derive its corresponding fractional Klein-Kramers equation, and explicitly investigate the diffusive behavior of the stochastic process described by this Langevin system for two cases of friction factor $\gamma=0$ and $\gamma\neq0$. Then we discuss the relations between CTRWs and Langevin system with different subordinators; the Langevin system with $\alpha$-dependent subordinator ($1<\alpha<2$) is presented in section \ref{Sec4} and another kind of Langevin system in section \ref{Sec5}. We give the method of generating $\alpha$-dependent subordinator with $1<\alpha<2$ and thus the subordinated processes for numerical simulations in section \ref{Sec6}. A summary of the key results is made in section \ref{Sec7}. In the appendices some mathematical details are collected.

\section{PDFs of $\alpha$-dependent subordinator ($1<\alpha<2$) and its inverse subordinator}\label{Sec2}

%\subsection{Subordinator}
A subordinator is a one-dimensional L\'{e}vy process that is non-decreasing (a.s.) \cite{Applebaum:2009}. Let $t(s)$ be a subordinator. Then the Laplace transform of its probability density function (PDF) is
\begin{equation}
  \hat{g}(\lambda,s):=\langle {\rm e}^{-\lambda t(s)}\rangle={\rm e}^{-s\Phi(\lambda)}.
\end{equation}
Here the bracket $\langle\cdots\rangle$ denotes the statistical average over stochastic realizations.
The Laplace exponent $\Phi(\lambda)$ takes the form \cite{Applebaum:2009}
\begin{equation*}
  \Phi(\lambda)=b\lambda+\int_0^\infty (1-{\rm e}^{-\lambda y}) \nu({\rm d}y),
\end{equation*}
where the drift $b\geq 0$ and the L\'{e}vy measure $\nu$ satisfies the additional requirements
\begin{equation}
  \nu(-\infty,0)=0 \quad {\rm and} \quad 
  \int_0^\infty \min\{y,1\}\nu({\rm d}y)<\infty.
\end{equation}
We call the pair $(b,\nu)$ the characteristics of the subordinator $t(s)$. If it is taken to be $b=0$ and
\begin{equation*}
  \nu({\rm d}y)=\frac{\alpha}{\Gamma(1-\alpha)}\frac{{\rm d}y}{y^{1+\alpha}},
\end{equation*}
with $0<\alpha<1$, then $\Phi(\lambda)=\lambda^\alpha$ is the Laplace exponent of the one-sided $\alpha$-dependent subordinator for $0<\alpha<1$, which has been fully discussed in Langevin systems \cite{Fogedby:1994,EuleFriedrichJenkoKleinhans:2007,CairoliBaule2:2015,ChenWangDeng:2018}. Here, we would like to specify the pair $(b,\nu)$ to form a subordinator for $1<\alpha<2$. For this purpose, considering the requirements of $\nu$, we take $b=0$ and
\begin{equation}\label{nu12}
  \nu({\rm d}y)=\frac{\alpha}{\tau_0}\frac{{\rm d}y}{(1+y/\tau_0)^{1+\alpha}},
\end{equation}
where $\tau_0$ is the characteristic time. Then its Laplace exponent reads
\begin{equation}\label{Phi}
  \Phi(\lambda)\simeq \mu_1\lambda-\mu_\alpha \lambda^\alpha,
\end{equation}
as $\lambda\rightarrow0$, where $\mu_1=\tau_0/(\alpha-1)>0$ and $\mu_\alpha=-\tau_0\Gamma(1-\alpha)>0$. The asymptotic behavior $\lambda\rightarrow0$ in Laplace space corresponds to $t\rightarrow\infty$ in the time domain, which is that people always pay attention to in physical experiments.

The two-point PDF of the subordinator $t(s)$ can be expressed as
\begin{equation*}
  g(t_2,s_2;t_1,s_1)= \langle \delta(t_2-t(s_2)) \delta(t_1-t(s_1))\rangle.
\end{equation*}
Its corresponding Laplace transform can be directly derived due to the Markovian character of this process. If $s_1<s_2$, considering the stationary and independent increments of the L\'{e}vy process, the characteristic function of $g(t_2,s_2;t_1,s_1)$ is \cite{BauleFriedrich:2005}
\begin{eqnarray}\label{g2lams}
    \hat{g}(\lambda_2,s_2;\lambda_1,s_1)&=\langle {\rm e}^{-\lambda_2t(s_2)-\lambda_1t(s_1)}\rangle  \nonumber \\
    &=\langle {\rm e}^{-\lambda_2(t(s_2)-t(s_1))}{\rm e}^{-(\lambda_2+\lambda_1)t(s_1)}\rangle \nonumber \\
    &={\rm e}^{-(s_2-s_1)\Phi(\lambda_2)}\,{\rm e}^{-s_1\Phi(\lambda_1+\lambda_2)}.
\end{eqnarray}
As for general $s_1$ and $s_2$, we usually write the characteristic function as
\begin{eqnarray*}
    \hat{g}(\lambda_2,s_2;\lambda_1,s_1)
    =&\Theta(s_2-s_1)\,{\rm e}^{-(s_2-s_1)\Phi(\lambda_2)}\,{\rm e}^{-s_1\Phi(\lambda_1+\lambda_2)}  \\
    &+\Theta(s_1-s_2)\,{\rm e}^{-(s_1-s_2)\Phi(\lambda_1)}\,{\rm e}^{-s_2\Phi(\lambda_1+\lambda_2)},
\end{eqnarray*}
where $\Theta(x)$ denotes the Heaviside step function: $\Theta(x)=1$ for $x>0$, $\Theta(x)=0$ for $x<0$, and $\Theta(x=0)=1/2$.

By the same approach, $n$-point joint PDF for the subordinator $t(s)$ can also be obtained. The multiple-point PDFs of (inverse) subordinator have been considered in the pioneering works \cite{BauleFriedrich:2005,BauleFriedrich:2007}. Here, we just extend the results to more general subordinator, e.g., the $\alpha$-dependent subordinator $t(s)$ with $1<\alpha<2$.

The first-passage time of a subordinator $\{t(s),s\geq 0\}$ is called inverse subordinator $\{s(t),t\geq 0\}$ \cite{KumarVellaisamy:2015,AlrawashdehKellyMeerschaertScheffler:2017}, defined as
\begin{equation}\label{inverse_st}
s(t)=\inf_{s>0}\{s:t(s)>t\}.
\end{equation}
Denote the (multiple-point) PDF of inverse subordinator $s(t)$ as
\begin{eqnarray*}
    h(s,t)=\langle \delta(s-s(t))\rangle, \\[3pt]
    h(s_2,t_2;s_1,t_1)= \langle \delta(s_2-s(t_2))\, \delta(s_1-s(t_1))\rangle.
\end{eqnarray*}
The specific expressions of PDF $h$ of inverse subordinator $s(t)$ can be derived through the intimate links with subordinator $t(s)$ \cite{BauleFriedrich:2005}:
\begin{eqnarray}
  \label{linkst1}  \langle \Theta(s-s(t))\rangle = 1-\langle \Theta(t-t(s))\rangle,  \\[3pt]
 \label{linkst2}  \fl \langle \Theta(s_2-s(t_2))\Theta(s_1-s(t_1))\rangle
    = 1-\langle\Theta(t_2-t(s_2))\rangle-\langle\Theta(t_1-t(s_1))\rangle \\
    \qquad\qquad\qquad\quad +\langle\Theta(t_2-t(s_2))\Theta(t_1-t(s_1))\rangle. \nonumber
\end{eqnarray}
Considering the formula that ${\rm d}\Theta(x)/{\rm d}x=\delta(x)$, taking the partial derivatives of $s$ or $s_1,s_2$ in (\ref{linkst1}) and (\ref{linkst2}), together with Laplace transform ($t\rightarrow\lambda,t_1\rightarrow\lambda_1,t_2\rightarrow\lambda_2$), we obtain the PDF of $h$:
\begin{equation}\label{h_1point}
  \hat{h}(s,\lambda)= -\frac{\partial}{\partial s} \frac{1}{\lambda}\, \hat{g}(\lambda,s)
                    = \frac{\Phi(\lambda)}{\lambda}\,{\rm e}^{-s\Phi(\lambda)}
\end{equation}
and
\begin{eqnarray}\label{h_2point}
  \fl  \hat{h}(s_2,\lambda_2;s_1,\lambda_1)
    =\frac{\partial^2}{\partial s_1\partial s_2} \frac{1}{\lambda_1\lambda_2}\,\hat{g}(\lambda_2,s_2;\lambda_1,s_1) \nonumber\\
    ~=\delta(s_2-s_1)\frac{\Phi(\lambda_1)+\Phi(\lambda_2)-\Phi(\lambda_1+\lambda_2)}{\lambda_1\lambda_2}\,{\rm e}^{-s_1\Phi(\lambda_1+\lambda_2)} \nonumber\\
    ~~~+\Theta(s_2-s_1)\frac{\Phi(\lambda_2)(\Phi(\lambda_1+\lambda_2)-\Phi(\lambda_2))}{\lambda_1\lambda_2}\,{\rm e}^{-s_1\Phi(\lambda_1+\lambda_2)} {\rm e}^{-(s_2-s_1)\Phi(\lambda_2)} \nonumber\\
    ~~~+\Theta(s_1-s_2)\frac{\Phi(\lambda_1)(\Phi(\lambda_1+\lambda_2)-\Phi(\lambda_1))}{\lambda_1\lambda_2} {\rm e}^{-s_2\Phi(\lambda_1+\lambda_2)}{\rm e}^{-(s_1-s_2)\Phi(\lambda_1)}.
\end{eqnarray}
Note that the PDFs in (\ref{h_1point}) and (\ref{h_2point}) are both normalized, i.e., $\int_0^\infty {\rm d}s \hat{h}(s,\lambda)=\lambda^{-1}$ and
$\int_0^\infty\!\!\int_0^\infty {\rm d}s_1{\rm d}s_2 \hat{h}(s_2,\lambda_2;s_1,\lambda_1)=\lambda_1^{-1}\lambda_2^{-1}$.
These PDFs of inverse subordinator play an important role in bridging the PDFs of the subordinated processes and original processes in a Langevin system.

The $\alpha_0$-stable subordinator for $0<\alpha_0<1$ is commonly used in Langevin system to describe subdiffusion in \cite{Fogedby:1994,MetzlerKlafter2:2000} or superdiffusion in \cite{FriedrichJenkoBauleEule:2006,EuleZaburdaevFriedrichGeisel:2012}. More kinds of subordinators (e.g., tempered stable, gamma, inverse Gaussian, and inverse inverse Gaussian subordinators) are considered in \cite{GajdaMagdziarz:2011,KumarWylomanskaPoloczanskiSundar:2017,WylomanskaKumarPoloczanskiVellaisamy:2016,KumarWylomanskaGajda:2017}. The (two-point) PDFs of inverse subordinator $s(t)$ in (\ref{h_1point}) and (\ref{h_2point}) can be directly applied to other inverse subordinators for a specific $\Phi(\lambda)$.

\section{Model}\label{Sec3}

We consider the following set of Langevin equations:
\begin{eqnarray}\label{model}
    \frac{{\rm d}}{{\rm d} t}x(t)=v(t), ~~~~~
    \frac{{\rm d}}{{\rm d} s}v(s)=-\gamma v(s) +\xi(s), ~~~~~
    \frac{{\rm d}}{{\rm d} s}t(s)= \eta(s),
\end{eqnarray}
where $\gamma$ is the friction coefficient, $\xi(s)$ is the Gaussian white noise satisfying $\langle\xi(s_1)\xi(s_2)\rangle=2D_v\delta(s_1-s_2)$, and $t(s)$ is the $\alpha$-dependent subordinator ($1<\alpha<2$) with Laplace exponent $\Phi(\lambda)\simeq \mu_1\lambda-\mu_\alpha \lambda^\alpha$ introduced in section \ref{Sec2}.
This model will be investigated in three aspects in the following three subsections: firstly, we derive the Klein-Kramers equation based on Feynman-Kac equation; then we discuss the diffusive behavior of $x(t)$ in two cases of $\gamma=0$ and $\gamma\neq0$.

\subsection{Fractional Klein-Kramers equation}
The fractional Klein-Kramers equation corresponding to (\ref{model}) can be directly obtained from the forward Feynman-Kac equation in \cite{CairoliBaule:2017,WangChenDeng:2018} (see also \cite{CairoliBaule:2015}), since $x(t)=\int_0^tv(t'){\rm d}t'$ could be interpreted as a functional of $v(t)$. The only difference with the one in \cite{CairoliBaule:2017} is that a new subordinator $t(s)$ with $1<\alpha<2$ is considered here. Fortunately, the method in \cite{CairoliBaule:2017,WangChenDeng:2018} can be applied to any  subordinator with Laplace exponent $\Phi(\lambda)$, which only makes a difference in fractional substantial derivative operator proposed by \cite{FriedrichJenkoBauleEule:2006}. In \cite{FriedrichJenkoBauleEule:2006,FriedrichJenkoBauleEule2:2006}, $t(s)$ is an $\alpha_0$-stable subordinator ($0<\alpha_0<1$) and the corresponding Laplace exponent is $\Phi_0(\lambda)=\lambda^{\alpha_0}$. In this case, the fractional Klein-Kramers equation governing the joint PDF $p(x,v,t)$ of position $x$ and velocity $v$ at time $t$ is
\begin{equation}\label{FKKE0}
  \left[\frac{\partial}{\partial t}+v\frac{\partial}{\partial x}\right] p(x,v,t) = \mathcal{L}_{\rm{FP}}\mathcal{D}_t^{1-\alpha_0}p(x,v,t),
\end{equation}
where $\mathcal{L}_{\rm{FP}}$ is the Fokker-Planck collision operator
\begin{equation*}
  \mathcal{L}_{\rm{FP}}= \gamma\frac{\partial}{\partial v}v+D_v\frac{\partial^2}{\partial v^2}
\end{equation*}
and $\mathcal{D}_t^{1-\alpha_0}$ is the fractional substantial derivative operator defined as \cite{FriedrichJenkoBauleEule:2006}
\begin{eqnarray*}
    \mathcal{D}_t^{1-\alpha_0} p(x,v,t)
    =\frac{1}{\Gamma(\alpha_0)}\left[\frac{\partial}{\partial t}+v\frac{\partial}{\partial x}\right]
    \int_0^t {\rm d}t'~ \frac{{\rm e}^{-(t-t')v\frac{\partial}{\partial x}}}{(t-t')^{1-\alpha_0}} p(x,v,t).
\end{eqnarray*}
Note that $\mathcal{D}_t^{1-\alpha_0}$ in (\ref{FKKE0}) comes from the inverse Fourier-Laplace transform ($\rho \rightarrow x,\lambda\rightarrow t$) of the symbol
\begin{equation*}
  \frac{\lambda+{\rm i}\rho v}{\Phi_0(\lambda+{\rm i}\rho v)}=(\lambda+{\rm i}\rho v)^{1-\alpha_0}.
\end{equation*}
With the new $\Phi(\lambda)$ in (\ref{Phi}) for the case of $1<\alpha<2$, we have
\begin{eqnarray*}
    \frac{\lambda+{\rm i}\rho v}{\Phi(\lambda+{\rm i}\rho v)}
    =\frac{1}{\mu_1-\mu_\alpha(\lambda+{\rm i}\rho v)^{\alpha-1}}
    \simeq \frac{1}{\mu_1}+\frac{\mu_\alpha}{\mu_1^2}(\lambda+{\rm i}\rho v)^{\alpha-1},
\end{eqnarray*}
as $\lambda\rightarrow0$ and $\rho \rightarrow0$.
Taking the inverse Fourier-Laplace transform, we get the operator
\begin{equation*}
  \tilde{\mathcal{D}}_t^{\alpha-1}:=\frac{1}{\mu_1}+\frac{\mu_\alpha}{\mu_1^2} \mathcal{D}_t^{\alpha-1}
\end{equation*}
and obtain the fractional Klein-Kramers equation in the case of $1<\alpha<2$
\begin{equation}\label{FKKE1}
  \left[\frac{\partial}{\partial t}+v\frac{\partial}{\partial x}\right] p(x,v,t) = \mathcal{L}_{\rm{FP}}\tilde{\mathcal{D}}_t^{\alpha-1}p(x,v,t).
\end{equation}

Integrating over the position $x$, or making the Fourier transform $(x\rightarrow \rho )$ together with letting $\rho =0$, the fractional equation governing the PDF of velocity $v$
\begin{equation}\label{FKKEv1}
  \frac{\partial}{\partial t}p(v,t)=\mathcal{L}_{\rm{FP}} \left(\frac{1}{\mu_1}+\frac{\mu_\alpha}{\mu_1^2}D_t^{\alpha-1}\right)p(v,t),
\end{equation}
is obtained, where $D_t^{\alpha-1}$ is the fractional Riemann-Liouville derivative operator \cite{Podlubny:1999} with Laplace symbol $\lambda^{\alpha-1}$, defined as
\begin{equation}
      D_t^{\alpha-1} p(v,t)
    =\frac{1}{\Gamma(2-\alpha)}\frac{\partial}{\partial t} \int_0^t  {\rm d}t' ~ \frac{p(v,t')}{(t-t')^{\alpha-1}} \quad {\rm for} ~ 1<\alpha<2.
\end{equation}
The corresponding equation governing the PDF of positive $x$ cannot be easily obtained by the similar procedure, since $v$ is embedded into the fractional substantial derivative operator $\tilde{\mathcal{D}}_t^{\alpha-1}$, where the time $t$ and position $x$ are coupled with each other. Hence, it seems not easy to get the PDF $p(x,t)$ and the moments of position $x$ from the Fokker-Planck equation of $x$. Instead, we will calculate the moments straightly from the Langevin system (\ref{model}).

\subsection{Moments for the case $\gamma=0$}
In the case of $\gamma=0$, the Langevin system (\ref{model}) reduces to
\begin{equation}\label{gamma0}
  \frac{{\rm d}}{{\rm d}t}x(t)=v(t), ~~~~~\frac{{\rm d}}{{\rm d}s}v(s)=\xi(s),~~~~~
   \frac{{\rm d}}{{\rm d}s}t(s)= \eta(s),
\end{equation}
which shows that $v$ is a standard Brownian motion with respect to operation time $s$. Denote $v(s)$ as the velocity in operation time and
$v(t):=v(s(t))$ in physical time. For convenience, we assume that the initial conditions are $x_0=v_0=0$. So the odd-order moments of $v$ and $x$ are all zero. For the even-order moments of $v$ and $x$,
we can firstly calculate the correlation function $\langle v(s_1)v(s_2)\rangle$ of velocity in (\ref{gamma0}) as
\begin{eqnarray}\label{vv1}
    \langle v(s_2)v(s_1)\rangle
    =\int_0^{s_2}\!\!\!\!\int_0^{s_1} {\rm d}s_2'{\rm d}s_1' ~ \langle \xi(s_2')\xi(s_1')\rangle
    =2D_v\cdot \min\{s_1,s_2\}.
\end{eqnarray}
The corresponding correlation function of $v(t)$ in physical time $t$ is given by \cite{BauleFriedrich:2005,BauleFriedrich:2007}
\begin{equation}\label{vv2}
  \langle v(t_2)v(t_1)\rangle =\int_0^{\infty}\!\!\!\!\int_0^{\infty} {\rm d}s_2{\rm d}s_1~\langle v(s_2)v(s_1)\rangle h(s_2,t_2;s_1,t_1).
\end{equation}
This relationship is due to the fact that the process $v(s)$ and subordinator $t(s)$ are statistically independent.
For convenience, we always make the calculations in Laplace space and obtain
\begin{equation}\label{vv3}
  \langle \hat{v}(\lambda_2)\hat{v}(\lambda_1)\rangle =\int_0^{\infty}\!\!\!\!\int_0^{\infty} {\rm d}s_2{\rm d}s_1~\langle v(s_2)v(s_1)\rangle \hat{h}(s_2,\lambda_2;s_1,\lambda_1).
\end{equation}
Substituting (\ref{h_2point}) and (\ref{vv1}) into (\ref{vv3}) gives
\begin{equation*}
  \langle \hat{v}(\lambda_2)\hat{v}(\lambda_1)\rangle \simeq \frac{2D_v}{\mu_1} \frac{1}{(\lambda_1+\lambda_2)\lambda_1\lambda_2},
\end{equation*}
as $\lambda_1,\lambda_2\rightarrow0$.
Taking the inverse Laplace transform provides the second moment of $v(t)$:
\begin{equation}\label{vmo2}
  \langle v^2(t)\rangle= \frac{2D_v}{\mu_1} t.
\end{equation}
Using the relation that ${\rm d}x(t)/{\rm d}t=v(t)$ in (\ref{gamma0}), we get the correlation function of $x(t)$:
\begin{equation*}
  \langle x(t_2)x(t_1)\rangle= \int_0^{t_2}\!\!\!\!\int_0^{t_1} {\rm d}t_2'{\rm d}t_1' ~ \langle v(t_2')v(t_1')\rangle,
\end{equation*}
and its expression in Laplace space:
\begin{eqnarray}\label{x2v2}
    \langle \hat{x}(\lambda_2)\hat{x}(\lambda_1)\rangle =  \frac{1}{\lambda_2\lambda_1} \langle \hat{v}(\lambda_2)\hat{v}(\lambda_1)\rangle
    \simeq \frac{2D_v}{\mu_1} \frac{1}{(\lambda_1+\lambda_2)\lambda_1^2\lambda_2^2},
\end{eqnarray}
after taking the inverse Laplace transform, which results in the second moment of $x(t)$:
\begin{equation}\label{xmo2}
  \langle x^2(t)\rangle = \frac{2D_v}{3\mu_1} t^3.
\end{equation}

Next, we calculate the fourth moment of $x(t)$. Similarly, the four-point correlation function of $v(t)$ should be presented firstly. In operation time $s$,
\begin{eqnarray*}
    \langle v(s_4)v(s_3)v(s_2)v(s_1)\rangle
    =\int_0^{s_4}\!\!\!\!\int_0^{s_3}\!\!\!\!\int_0^{s_2}\!\!\!\!\int_0^{s_1} {\rm d}s_4'{\rm d}s_3'{\rm d}s_2'{\rm d}s_1' ~\langle \xi(s_4')\xi(s_3')\xi(s_2')\xi(s_1')\rangle,
\end{eqnarray*}
where the integrand equals to \cite{Risken:1989}
\begin{eqnarray*}
    4D_v^2[\delta(s_1'-s_2')\delta(s_3'-s_4')+\delta(s_1'-s_3')\delta(s_2'-s_4')
        +\delta(s_1'-s_4')\delta(s_2'-s_3')].
\end{eqnarray*}
For simplicity, we assume $s_1<s_2<s_3<s_4$, which leads to
\begin{equation}\label{v4s}
  \langle v(s_4)v(s_3)v(s_2)v(s_1)\rangle = 4D_v^2 (s_1s_3+2s_1s_2).
\end{equation}

Similarly to (\ref{vv2}), it seems that the four-point distribution $h$ of inverse subordinator $s(t)$ is needed to calculate $\langle v(t_4)v(t_3)v(t_2)v(t_1)\rangle$ in physical time $t$, which might be too complicated or even unavailable. But following (\ref{v4s}), it could be directly given as
\begin{equation}\label{v4t}
  \langle v(t_4)v(t_3)v(t_2)v(t_1)\rangle = 4D_v^2 (\langle s(t_1)s(t_3)\rangle+2\langle s(t_1)s(t_2)\rangle).
\end{equation}
The formula (\ref{v4t}) provides a shortcut and reduces the four-point distribution $h$ to two-point. But that (\ref{v4t}) holds has the preconditional hypothesis $s_1<s_2<s_3<s_4$.
We claim that (\ref{v4t}) is valid on the condition that $t_1<t_2<t_3<t_4$ and provide the detailed derivations in \ref{App1}. The techniques used will also work in other places.

% will be useful everywhere.

Two-point correlation function $\langle s(t_1)s(t_2)\rangle$ can be directly calculated using (\ref{h_2point}). After some lengthly calculations in Laplace space, we get
\begin{eqnarray}\label{correl_s2}
    \langle s(\lambda_1)s(\lambda_2)\rangle
    &= \int_0^\infty\!\!\int_0^\infty {\rm d}s_1{\rm d}s_2 \,s_1s_2\hat{h}(s_2,\lambda_2;s_1,\lambda_1) \nonumber\\
    &=  \frac{1}{\lambda_1\lambda_2\Phi(\lambda_1+\lambda_2)}\left(\frac{1}{\Phi(\lambda_1)}+\frac{1}{\Phi(\lambda_2)}\right)  \\[3pt]
    & \simeq  \mu_1^{-2} \, (\lambda_1\lambda_2)^{-2}, \nonumber
\end{eqnarray}
as $\lambda_1,\lambda_2\rightarrow0$, which implies that $\langle s(t_1)s(t_2)\rangle \simeq \mu_1^{-2} t_1t_2$ and thus
\begin{equation}\label{v4t2}
  \langle v(t_4)v(t_3)v(t_2)v(t_1)\rangle = \frac{4D_v^2}{\mu_1^2} ( t_1t_3+2t_1t_2)
\end{equation}
for $t_1<t_2<t_3<t_4$. Letting $t_1=t_2=t_3=t_4=t$, we get
\begin{equation}\label{vmo4}
  \langle v^4(t)\rangle \simeq \frac{12D_v^2}{\mu_1^2}t^2.
\end{equation}
For the fourth moment of $x(t)$, it can be written as
\begin{eqnarray*}
  \langle x^4(t)\rangle
  &=\int_0^t\!\!\int_0^t\!\!\int_0^t\!\!\int_0^t\!\!{\rm d}t_4{\rm d}t_3{\rm d}t_2{\rm d}t_1\langle v(t_4)v(t_3)v(t_2)v(t_1)\rangle \\
  &=4!\int_0^t{\rm d}t_4\int_0^{t_4}{\rm d}t_3\int_0^{t_3}{\rm d}t_2\int_0^{t_2}{\rm d}t_1\langle v(t_4)v(t_3)v(t_2)v(t_1)\rangle.
\end{eqnarray*}
Substituting the result (\ref{v4t2}) into above formula, we obtain
\begin{equation}\label{xmo4}
   \langle x^4(t)\rangle \simeq \frac{4D_v^2}{3\mu_1^2}t^6.
\end{equation}

\begin{figure}
  \centering
  \includegraphics[scale=0.6]{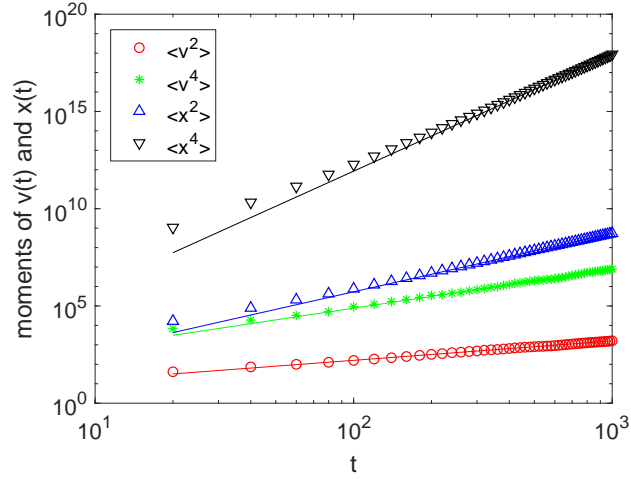}\\
  \caption{Second and fourth moments of velocity $v(t)$ and position $x(t)$ versus physical time $t$. 1000 trajectories are used with parameters: $T=1000$, $D_v=1$, $\gamma=0$, $\alpha=1.8$, and $\tau_0=1$. The solid lines denote the theoretical results for long times while the markers the simulation results.}\label{fig1}
\end{figure}

The low-order moments of velocity $v(t)$ and position $x(t)$ have been obtained in (\ref{vmo2}), (\ref{xmo2}), (\ref{vmo4}) and (\ref{xmo4}), with their numerical simulations presented in figure \ref{fig1}. Note that these results only differ with the Langevin system without subordinator $t(s)$ by a prefactor. This maybe understandable since the mean value of subordinator $t(s)$ exists. In this sense, our subordinator $t(s)$ just changes the time scale in a linear way by the parameter $\mu_1$ for long times. But this is just a special case for $\gamma=0$. In the following section, we consider $\gamma\neq0$ and show the non-trivial moments.

\subsection{Moments for the case $\gamma\neq0$}
In the case of $\gamma\neq0$, the velocity $v(s)$ in (\ref{model}) is not a Brownian motion, but an Ornstein-Uhlenbeck process \cite{Risken:1989}, which ensures a steady state of the diffusivity dynamics with respect to velocity for long times. The velocity process $v(s)$ can be analytically expressed as
\begin{equation*}
  v(s)=\int_0^s {\rm d}s'~ \xi(s'){\rm e}^{-\gamma(s-s')}+ v_0{\rm e}^{-\gamma s},
\end{equation*}
which implies that the mean of $v(s)$ is $\langle v(s)\rangle=v_0 {\rm e}^{-\gamma s}$, tending to zero for long times, and the correlation function of $v(s)$ is
\begin{eqnarray}\label{vs_2point}
    \langle v(s_1)v(s_2)\rangle
      = \frac{D_v}{\gamma}({\rm e}^{-\gamma|s_1-s_2|}-{\rm e}^{-\gamma(s_1+s_2)})
       + v_0^2 {\rm e}^{-\gamma(s_1+s_2)} .
\end{eqnarray}
Then the second moment of $v$ in operation time $s$ reads
\begin{equation*}
  \langle v^2(s)\rangle = \frac{D_v}{\gamma}+\left(v_0^2-\frac{D_v}{\gamma}\right) {\rm e}^{-2\gamma s}.
\end{equation*}
The second moment of $v$ in physical time $t$ can be obtained by using the relation \cite{BauleFriedrich:2005}
\begin{equation*}
  p(v,t)=\int_0^\infty {\rm d}s~ p_0(v,s)h(s,t),
\end{equation*}
where $p(v,t)$ and $p_0(v,s)$ denote the PDFs of $v(t)$ and $v(s)$,  respectively. Multiplying $v^2$ on both sides and integrating over $v$, together with Laplace transform and (\ref{h_1point}), we get
\begin{eqnarray*}
    \langle \hat{v}^2(\lambda)\rangle &= \int_0^\infty {\rm d}s~ \langle v^2(s)\rangle \hat{h}(s,\lambda)
    \simeq \frac{D_v}{\gamma}\cdot\frac{1}{\lambda}+\left(v_0^2-\frac{D_v}{\gamma}\right) \frac{1}{\lambda+2\gamma/\mu_1},
\end{eqnarray*}
and thus
\begin{eqnarray}\label{vmo22}
    \langle v^2(t)\rangle &\simeq \frac{D_v}{\gamma}+\left(v_0^2-\frac{D_v}{\gamma}\right) {\rm e}^{-2\gamma t/\mu_1}
    \simeq \frac{D_v}{\gamma}
\end{eqnarray}
for long times.

\begin{figure}
  \centering
  \includegraphics[scale=0.6]{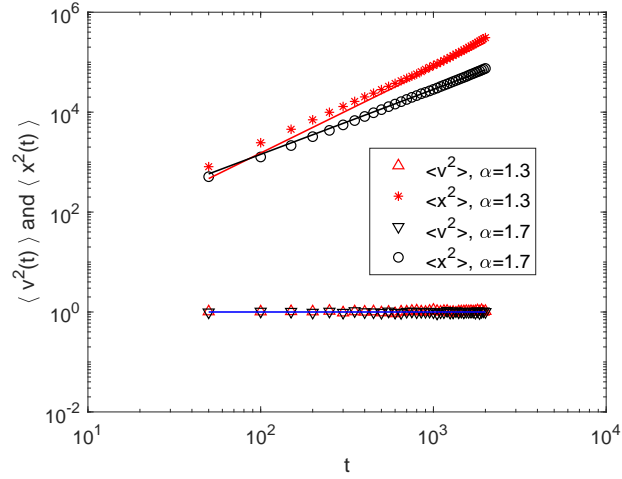}\\
  \caption{Second moments of velocity $v(t)$ and position $x(t)$ versus physical time $t$. 1000 trajectories are used with parameters: $T=2000$, $D_v=1$, $\gamma=1$, $\alpha=1.3,1.7$, and $\tau_0=1$. The solid lines denote the theoretical results for long times while the markers the simulation results.}\label{fig2}
\end{figure}

For the second moment of $x$, we resort to (\ref{h_2point}) and (\ref{vv3}) and obtain
\begin{eqnarray}\label{v2lam2}
    \langle \hat{v}(\lambda_1)\hat{v}(\lambda_2)\rangle &= \frac{D_v}{\gamma\lambda_1\lambda_2}\cdot \left(\frac{1}{\Phi(\lambda_1+\lambda_2)}-\frac{1}{\Phi(\lambda_1+\lambda_2)+2\gamma}\right)   \nonumber\\
   &~~\times \left(\frac{\Phi(\lambda_1)\Phi(\lambda_2)\Phi(\lambda_1+\lambda_2)+2\Phi(\lambda_1)\Phi(\lambda_2)\gamma}{[\Phi(\lambda_1)+\gamma][\Phi(\lambda_2)+\gamma]}\right.  \nonumber\\
     &~~\left.+\frac{[\Phi(\lambda_1)+\Phi(\lambda_2)-\Phi(\lambda_1+\lambda_2)]\gamma^2}{[\Phi(\lambda_1)+\gamma][\Phi(\lambda_2)+\gamma]} \right) \nonumber\\
   &\simeq \frac{D_v}{\gamma\lambda_1\lambda_2}\cdot \frac{\Phi(\lambda_1)+\Phi(\lambda_2)-\Phi(\lambda_1+\lambda_2)}{\Phi(\lambda_1+\lambda_2)}.
\end{eqnarray}
Considering $\Phi(\lambda)=\mu_1\lambda-\mu_\alpha\lambda^\alpha$, we have
\begin{equation}\label{v2lam22}
  \langle \hat{v}(\lambda_1)\hat{v}(\lambda_2)\rangle
  \simeq \frac{D_v\mu_\alpha}{\gamma\mu_1}\cdot \frac{(\lambda_1+\lambda_2)^\alpha-\lambda_1^\alpha-\lambda_2^\alpha}{\lambda_1\lambda_2(\lambda_1+\lambda_2)}
\end{equation}
and
\begin{eqnarray}\label{x2lam2}
  \langle \hat{x}(\lambda_1)\hat{x}(\lambda_2)\rangle
    \simeq \frac{D_v\mu_\alpha}{\gamma\mu_1} \cdot \frac{(\lambda_1+\lambda_2)^\alpha-\lambda_1^\alpha-\lambda_2^\alpha}{\lambda_1^2\lambda_2^2(\lambda_1+\lambda_2)}
\end{eqnarray}
as $\lambda_1,\lambda_2\rightarrow0$.
Inversing (\ref{v2lam22}) and (\ref{x2lam2}), the correlation function of $v(t)$ and $x(t)$ can be obtained (presented in \ref{App2}). Letting $t_1=t_2=t$ there, we get
\begin{equation}\label{xmo22}
  \langle x^2(t)\rangle \simeq \frac{D_v\mu_\alpha}{\gamma\mu_1}\frac{2\alpha-2}{\Gamma(4-\alpha)}\,t^{3-\alpha},
\end{equation}
for long times.
The simulation results of the second moments of $v(t)$ and $x(t)$ with $\alpha=1.3$ and $\alpha=1.7$ are shown in figure \ref{fig2}, which are consistent to the theoretical results in solid lines for long times.

It can be seen that the second moment of $x(t)$ depends on $\alpha$. This result is different from the case without the subordinator, in which \cite{Risken:1989}
\begin{equation*}
   \qquad \langle x^2(t)\rangle \propto t.
\end{equation*}
If we pay attention to the correlation function of $x(t)$, the main difference from the case of $\gamma=0$ in the previous subsection (\ref{x2v2}) is that $\gamma\neq0$ here makes the asymptotic expression of the correlation function of $v(t)$ (\ref{v2lam22}) depend on $\alpha$. More essential reasons about the difference the new subordinator brings in will be discussed in the next section.
By adding a harmonic potential on $v$ (i.e., $\gamma\neq0$) in (\ref{model}), the correlation function of $v(t)$ is obtained in (\ref{v2lam2}). This result can be extended to a system within an arbitrary confined potential $U(v)$, where the steady state on $v$ can be achieved. In this case, we denote the average of an observable $\mathcal{O}(v)$ on the Boltzmann distribution as
\begin{equation}
  \langle \mathcal{O}(v)\rangle_B = \frac{1}{N}\int_{-\infty}^{\infty}{\rm d}v\mathcal{O}(v)\exp[-U(v)/k_BT],
\end{equation}
where $N=\int_{-\infty}^{\infty}{\rm d}v\exp[-U(v)/k_BT]$ is the normalizing function, and $k_BT$ the thermal energy.  By imitating the method in \cite{BurovMetzlerBarkai:2010}, the correlation function of $v(t)$ in confined potential $U(v)$  can be presented in Laplace space as
\begin{eqnarray}\label{pentialU}
  \langle \hat{v}(\lambda_1)\hat{v}(\lambda_2)\rangle &=\frac{\Phi(\lambda_1)+\Phi(\lambda_2)-\Phi(\lambda_1+\lambda_2)}{\lambda_1\lambda_2\Phi(\lambda_1+\lambda_2)}
  \Big(\langle v^2\rangle_B-\langle v\rangle^2_B\Big) + \frac{\langle v\rangle^2_B}{\lambda_1\lambda_2},
\end{eqnarray}
which recovers (\ref{v2lam2}) when $\langle v^2\rangle_B=D_v/\gamma$ and $\langle v\rangle_B=0$ in model (\ref{model}).

\begin{figure}
\flushright
\begin{minipage}{0.35\linewidth}
  \centerline{\includegraphics[scale=0.45]{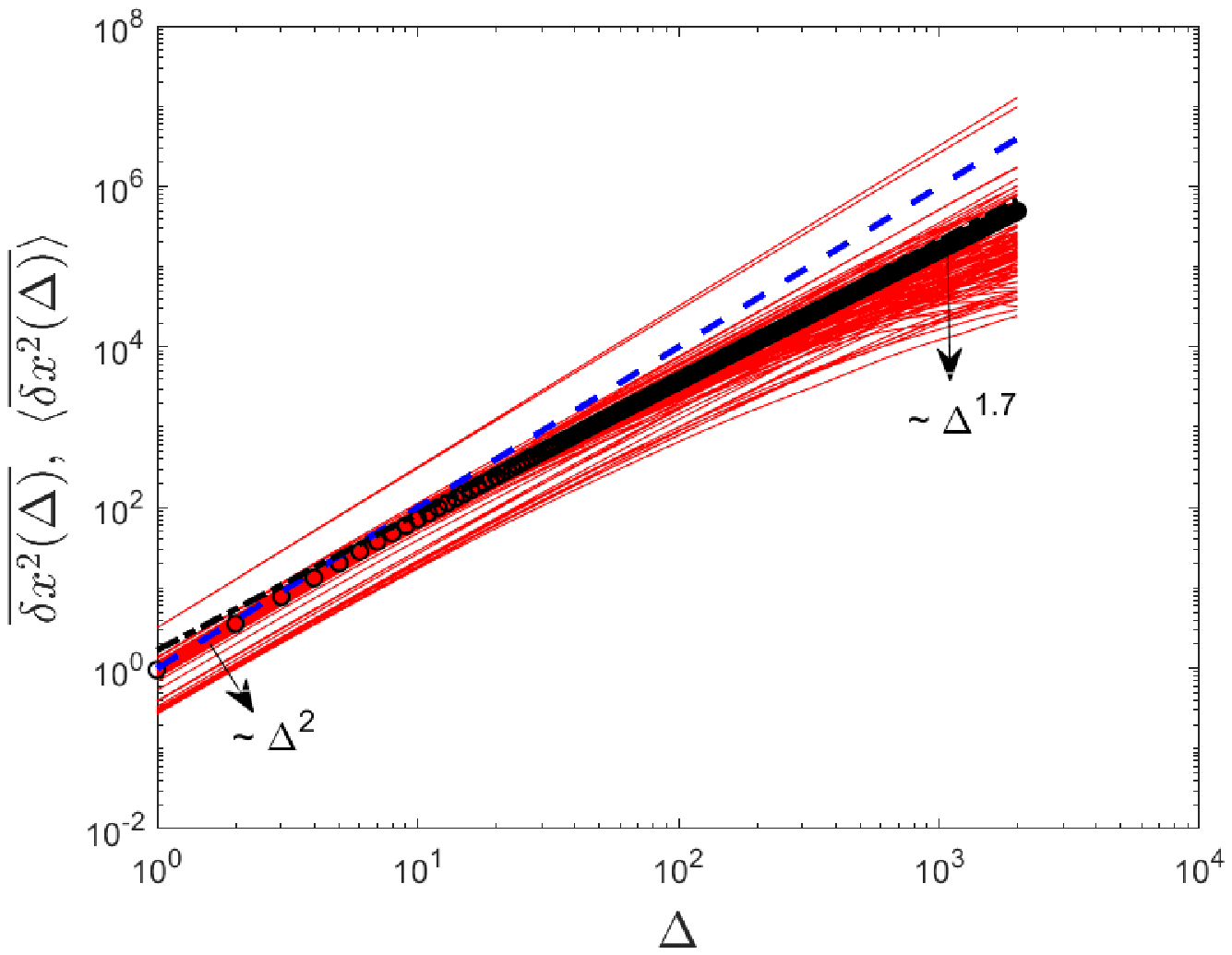}}
  \centerline{(a) $\alpha=1.3$}
\end{minipage}
%\hfill
\hspace{1.2cm}
%\vfill
\begin{minipage}{0.35\linewidth}
  \centerline{\includegraphics[scale=0.45]{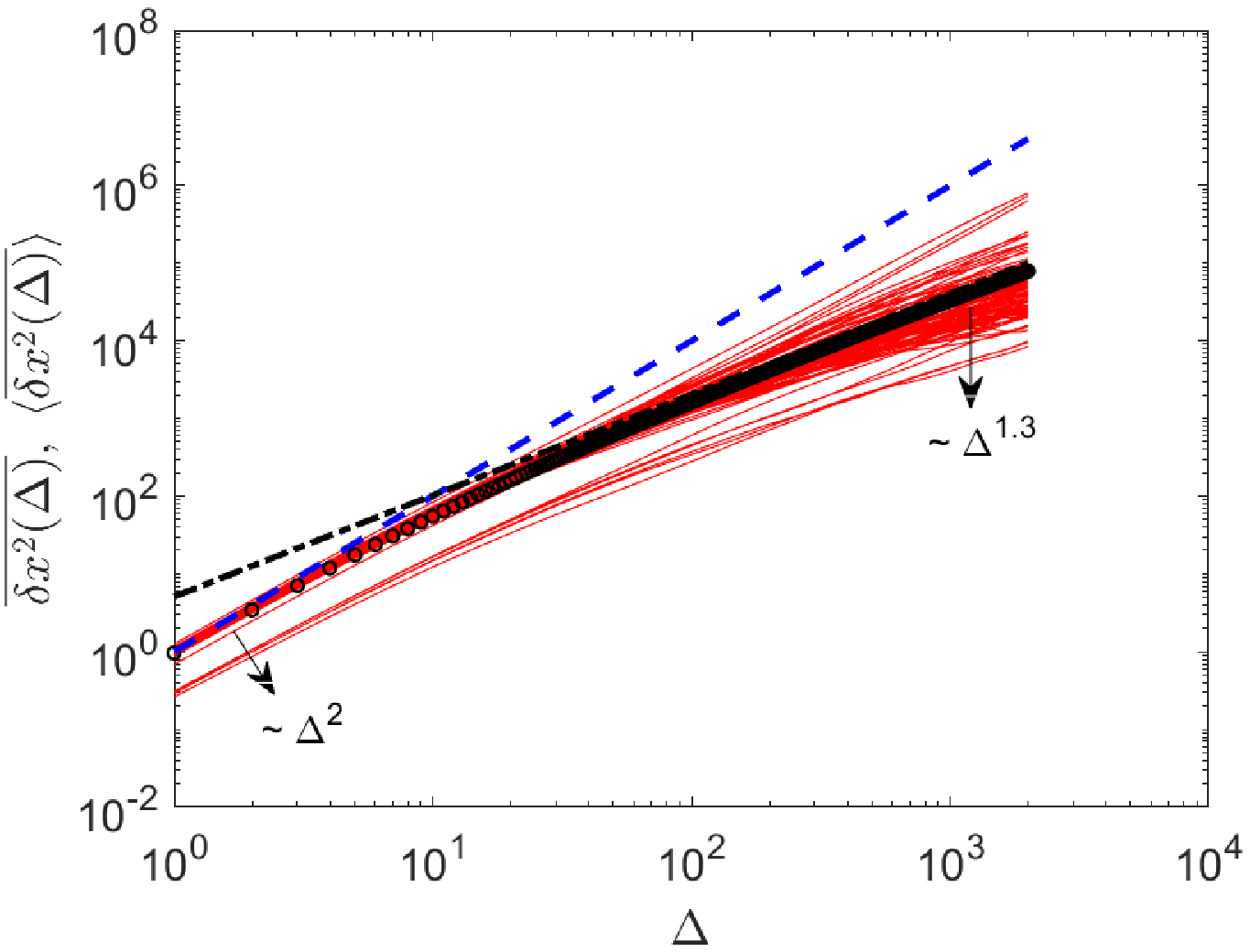}}
  \centerline{(b) $\alpha=1.7$}
\end{minipage}
  \caption{Time averaged MSDs $\overline{\delta x^2(\Delta)}$ for $T=2000,D_v=1,\gamma=1$, and $\tau_0=1$. The thin red lines show the results for individual time averaged trajectories of one hundred samples. The black circles denote the trajectory average $\langle\overline{\delta x^2(\Delta)}\rangle$; the black dash dot lines refer to the theoretical results (\ref{TAMSD_A}) $\propto \Delta^{3-\alpha}$ while the blue dash lines $\propto \Delta^{2}$. We observe that the regime of $\langle\overline{\delta x^2(\Delta)}\rangle$ changes from $\Delta^{2}$ to $\Delta^{3-\alpha}$, which looks more obvious in (b) for larger $\alpha$.}\label{fig4}
\end{figure}

When constructing single particle tracking experiments, the process $x(t)$ is evaluated in terms of the time averaged MSD, defined via
\begin{equation}
  \overline{\delta x^2(\Delta)}=\frac{1}{T-\Delta}\int_0^{T-\Delta}{\rm d}t[x(t+\Delta)-x(t)]^2,
\end{equation}
$\Delta$ denoting the lag time. Typically, $\overline{\delta x^2(\Delta)}$ is considered in the limit $\Delta\ll T$ to obtain good statistics. The correlation function of $x(t)$ in the integrand depends on the correlation function of $v(t)$ in (\ref{correlation_v}) that
\begin{equation}\label{correlation_vv}
  \langle v(t_1)v(t_2)\rangle= \frac{D_v\mu_\alpha}{\gamma\mu_1}\cdot \frac{(t_2-t_1)^{1-\alpha}-t_2^{1-\alpha}}{\Gamma(2-\alpha)}.
\end{equation}
Alternatively, the time averaged MSD can also be obtained from the corresponding time averaged velocity correlation function \cite{GodecMetzler:2013}
\begin{equation}\label{TAcorr_v}
    C_v(\tau)=\frac{1}{T-\tau}\int_0^{T-\tau}{\rm d}t\, v(t)v(t+\tau),
\end{equation}
and the Green-Kubo formula \cite{Kubo:1966}
\begin{equation}\label{GKform}
  \overline{\delta x^2(\Delta)}=2\int_0^\Delta {\rm d}\tau(\Delta-\tau)C_v(\tau).
\end{equation}
Substituting the correlation function of $v(t)$ in (\ref{correlation_vv}) into (\ref{TAcorr_v}) and (\ref{GKform}),
we obtain the mean of the time averaged MSD
\begin{equation}\label{TAMSD_A}
  \langle\overline{\delta x^2(\Delta)}\rangle \simeq\frac{D_v\mu_\alpha}{\gamma\mu_1}\cdot \frac{2\Delta^{3-\alpha}}{\Gamma(4-\alpha)}
\end{equation}
for $\Delta<\!\!<T$. This result has been simulated in figure \ref{fig4} with different $\alpha$.
It shows that the averaged quantity $\langle\overline{\delta x^2(\Delta)}\rangle$ experiences $\Delta^2$ in short time and $\Delta^{3-\alpha}$ in long time, consistent to the result in \cite{GodecMetzler:2013} about the L\'{e}vy walk.
Define the ergodicity-breaking parameter as the ratio of time versus ensemble averaged MSD. Combining (\ref{xmo22}) and (\ref{TAMSD_A}) shows
\begin{equation}
    \mathcal{EB}= \frac{\langle\overline{\delta x^2(\Delta)}\rangle}{\langle x^2(\Delta)\rangle}=\frac{1}{\alpha-1},
\end{equation}
which implies that the MSD is ultraweak ergodicity breaking, consistent to the results of L\'{e}vy walk in \cite{GodecMetzler:2013,FroembergBarkai:2013}. From the MSD of $x(t)$ (\ref{xmo22}) and the time averaged MSD (\ref{TAMSD_A}), we deem that the model (\ref{model}) with $\gamma\neq0$ describes the motion like L\'{e}vy walk. Especially for $\alpha$-dependent subordinator with $1<\alpha<2$, it corresponds to the L\'{e}vy walk of sub-ballistic superdiffusion regime.

\section{Relation with CTRWs and L\'{e}vy walk}\label{Sec4}
In CTRWs, the motion of a particle is described by consecutive random waiting times between random jumps. The particle may undergo normal  or anomalous diffusion, depending on whether the distributions are heavy-tailed or not. One special case is L\'{e}vy flight \cite{ChechkinGoncharKlafterMetzlerTanatarov:2004,BrockmannGeisel:2003,DybiecGudowska:2009}, where the waiting times have finite mean value but the jump lengths have infinite second moment.
The possible disadvantages of L\'{e}vy flight are the diverging mean square displacement and the infinite velocity, which may be lack of physical meaning for a particle with finite mass.
But L\'{e}vy walk avoids these drawbacks, where waiting time and jump length are coupled with each other. The standard L\'{e}vy walk says a particle moves ballistically for a random time and then randomly changes direction but keeps the same magnitude of velocity \cite{ZaburdaevDenisovKlafter:2015}. Therefore, in L\'{e}vy walk, much time penalizes a large jump and this balances the velocity to be finite.

Fogedby \cite{Fogedby:1994} proposed the coupled Langevin equation (\ref{model2}) to describe the process in CTRWs, where $\xi(s)$ and $\eta(s)$ are independent with each other, characterizing the jump lengths and waiting times, respectively. Commonly, $\eta(s)$ is taken to be one-sided $\alpha_0$-stable ($0<\alpha_0<1$) for describing the heavy-tailed waiting times distribution, and $\xi(s)$ might be $\beta$-stable ($0<\beta<2$) for characterizing heavy-tailed jump lengths distribution in CTRWs.
But for L\'{e}vy walk, its corresponding Langevin picture should be presented like (\ref{model}), where the derivative of position $x$ with respect to physical time $t$ is velocity $v$ and the subordinator $t(s)$ characterizes the distribution of duration of each flight. The second equation in (\ref{model}) gives the distribution of velocity $v$. One special case that $\eta(s)$ is a one-sided $\alpha_0$-stable distribution ($0<\alpha_0<1$) and
$$v(s)=\gamma^{-1}\xi_d(s)$$
has been pointed out in \cite{EuleZaburdaevFriedrichGeisel:2012}, where $\xi_d(s)$ is a dichotomous noise source, i.e., a random sequence of the values $-1$ and $1$. It is just a one-to-one correspondence to the standard L\'{e}vy walk with the exponent of waiting time distribution less than $1$. In general, the distribution of velocity $v$ could be various, such as, Gaussian distribution, exponential distribution, and uniform distribution \cite{RebenshtokDenisovHanggiBarkai:2014,RebenshtokDenisovHanggiBarkai2:2014}. In more general cases, velocity $v$ may be 
fluctuant due to a random force \cite{KaratsasShreve:1977} and thus its distribution becomes time-dependent. All in all, velocity $v$ can be described by a Langevin equation, i.e., the second equation of (\ref{model}). The nonzero constant $\gamma$ makes sure a steady state of velocity $v$ could be reached for long times, analogously to the finite moments of $v$ in L\'{e}vy walk.
In a word, the overdamped Langevin equation with a subordinator (\ref{model2}) corresponds to CTRWs, while the weakly damped Langevin equation coupled with a subordinator (\ref{model}) corresponds to L\'{e}vy walks. More generally, formula (\ref{pentialU}) in some sense implies that L\'{e}vy walk can also be modeled by arbitrary confined potential $U(v)$ in velocity not only harmonic potential; the harmonic petential together with Gaussian white noise may be the simplest choice. For an asymmetric confined potential $U(v)$, the biased L\'{e}vy walk together with the correlation function of $v(t)$ can also be obtained.

For the coupled Langevin equation (\ref{model}), one-sided $\alpha_0$-stable subordinator ($0<\alpha_0<1$) has been considered in \cite{FriedrichJenkoBauleEule:2006,FriedrichJenkoBauleEule2:2006}, where the second moment of position $x$ is
\begin{equation}\label{alpha0_1}
  \langle x^2(t)\rangle \propto t^{2+\alpha_0}  \quad \rm{with}~\gamma=0,
\end{equation}
and
\begin{equation}\label{alpha0_2}
  \langle x^2(t)\rangle \propto t^2  \quad \rm{with}~\gamma\neq0.
\end{equation}
The result with $\gamma\neq0$ is consistent to the standard L\'{e}vy walk in ballistic regime. Furthermore, we extend the subordinator to be $\alpha$-dependent ($1<\alpha<2$) and obtain the sub-ballistic superdiffusion regime (\ref{xmo22}) consistent to the corresponding L\'{e}vy walk. These two cases confirm the statement that the Langevin system (\ref{model}) models the L\'{e}vy walk in long times.
This subordinator of $1<\alpha<2$ has never been considered, but it is important to give rise to strong anomalous diffusion in this system, where the moments $\langle |x(t)|^q\rangle$ exhibit different diffusion scales for different ranges of $q$ \cite{RebenshtokDenisovHanggiBarkai:2014,RebenshtokDenisovHanggiBarkai2:2014}.

We find an intriguing phenomenon that, compared to the second moments of position $x$ with $0<\alpha_0<1$ in (\ref{alpha0_1}) and (\ref{alpha0_2}),  the diffusive behavior in (\ref{xmo2}) and (\ref{xmo22}) with $1<\alpha<2$ is enhanced for $\gamma=0$ but suppressed for $\gamma\neq0$. The Langevin system (\ref{model}) with $\gamma=0$ or $\gamma\neq0$ are completely different models.
Since $\alpha$-dependent subordinator characterizes the heavy-tailed distribution of waiting times in CTRWs, it may yield longer waiting time for $0<\alpha<1$ than $1<\alpha<2$.
For $\gamma=0$, the subordinator suppresses the diffusion of velocity $v$ and thus $x$ due to the occasionally long waiting time, which implies the diffusion with $0<\alpha<1$ is suppressed more seriously. But for $\gamma\neq0$, velocity $v$ can reach a steady state for long time and the subordinator suppresses the rate of changing direction of particles and thus enhances the diffusion of displacement $x$, which results in a contrary result compared with $\gamma=0$.

It is worth to note that the $\alpha$-dependent subordinator does not always contribute to the strong anomalous diffusion phenomenon. Sometimes it makes a trivial result, like the case of $\gamma=0$ in (\ref{gamma0}), where the moments of $x(t)$ exhibit single diffusion scale.
Actually, the position $x(t)$ is a stochastic process with self-similarity, which can be briefly demonstrated. The $\alpha$-dependent subordinator $t(s)$ is $1/\tilde{\alpha}$ self-similar \cite{Applebaum:2009}, where $\tilde{\alpha}=\alpha$ for $0<\alpha<1$ and $\tilde{\alpha}=1$ for $1<\alpha<2$. And then the inverse subordinator $s(t)$ is $\tilde{\alpha}$ self-similar \cite{MagdziarzMetzlerSzczotkaZebrowski:2012}. Therefore, the coupled velocity process
\begin{eqnarray}\label{similarity}
    v(t):=v(s(t))=B(s(t))\overset{d}{=}B(t^{\tilde{\alpha}}s(1))
          \overset{d}{=}t^{\tilde\alpha/2}B(s(1))=t^{\tilde\alpha/2}v(1),
\end{eqnarray}
where $\overset{d}{=}$ denotes identical distribution.
Formula (\ref{similarity}) implies $v(t)$ is $\tilde\alpha/2$ self-similar and thus $x(t)$ is $\tilde\alpha/2+1$ self-similar. In this way,
\begin{equation}
  \langle |x(t)|^n\rangle \propto t^{n(\tilde\alpha/2+1)}.
\end{equation}
This single diffusion scale indicates that there is no strong anomalous diffusion. But for $\gamma\neq0$ or a more general nonlinear external force, which can be naturally added into the Langevin system, this subordinator might introduce a multiple diffusion scales and a different diffusion phenomenon.

\section{Comparison with another Langevin system}\label{Sec5}
Different from the Langevin system (\ref{model}), another kind of commonly considered coupled Langevin system is 
\begin{eqnarray}\label{model1}
    \frac{{\rm d}}{{\rm d}s}x(s)=v(s), ~~~~~
    \frac{{\rm d}}{{\rm d}s}v(s)=-\gamma v(s) +\xi(s), ~~~~~
    \frac{{\rm d}}{{\rm d}s}t(s)= \eta(s),
\end{eqnarray}
where position $x$ and velocity $v$ are both subordinated. If $t(s)$ is the $\alpha$-dependent subordinator ($0<\alpha<1$), its corresponding fractional Klein-Kramers equation is proposed in \cite{MetzlerKlafter2:2000}. Here we consider the case of $1<\alpha<2$, and the fractional Klein-Kramers equation will be different from (\ref{FKKE1}). Denote the joint PDF of position $x$ and velocity $v$ in operation time as $p_0(x,v,s)$ and the one in physical time $p(x,v,t)$. Then $p_0(x,v,s)$ solves the Klein-Kramers equation \cite{CoffeyKalmykovWaldron:2004}
\begin{equation*}
  \left[\frac{\partial}{\partial s}+v\frac{\partial}{\partial x}\right]p_0(x,v,s) = \mathcal{L}_{\mathrm{FP}}\,p_0(x,v,s).
\end{equation*}
Using the relation
\begin{equation*}
  p(x,v,t)=\int_0^{\infty}{\rm d}s ~p_0(x,v,s)h(s,t),
\end{equation*}
we have
\begin{eqnarray}\label{FKKE2}
    \frac{\partial}{\partial t}p(x,v,t) = \left[-v\frac{\partial}{\partial x} + \mathcal{L}_{\mathrm{FP}}\right]
    \left(\frac{1}{\mu_1}+\frac{\mu_\alpha}{\mu_1^2}D_t^{\alpha-1}\right)\,p(x,v,t).
\end{eqnarray}
This is the fractional Klein-Kramers equation governing the joint PDF of position-velocity of the Langevin system (\ref{model1}). Note that in this case, the Newton relation does not hold between $x(t)$ and $v(t)$ and Galilean invariance is violated \cite{MetzlerKlafter:2000,EuleFriedrich:2009}.

Integrating over the position $x$ on (\ref{FKKE2}), we get the same equation governing the PDF of velocity $v(t)$ as (\ref{FKKEv1}). This is reasonable since the only difference between the Langevin system (\ref{model}) and (\ref{model1}) is the position $x(t)$. But here, we can also derive the equation governing the PDF of position $x(t)$ by integrating (\ref{FKKE2}) over $\int {\rm d}v$ and $\int v{\rm d}v$, and combining the two resulted equations. With $\langle v^2(t)\rangle\simeq D_v/\gamma$ in (\ref{vmo22}) for the case of $\gamma\neq0$, this procedure yields the fractional diffusion equation of $p(x,t)$:
\begin{eqnarray}\label{FKKEx2}
  \fl  \frac{\partial^2}{\partial t^2}p(x,t)+\gamma\left(\frac{1}{\mu_1}\frac{\partial}{\partial t}+\frac{\mu_\alpha}{\mu_1^2}D_t^\alpha\right)p(x,t)
    =\frac{D_v}{\gamma}\frac{\partial^2}{\partial x^2}\left(\frac{1}{\mu_1}+\frac{\mu_\alpha}{\mu_1^2}D_t^{\alpha-1}\right)^2p(x,t),
\end{eqnarray}
which becomes, in the long time or high-friction limit,
\begin{equation}\label{FKKEx22}
  \frac{\partial}{\partial t}p(x,t) = \frac{D_v}{\gamma^2\mu_1}\frac{\partial^2}{\partial x^2}p(x,t).
\end{equation}
It can be seen that in the long time limit, the Langevin system (\ref{model1}) undergoes normal diffusion, with the odd-order moments vanishing and even-order moments as
\begin{equation}\label{model1_x2n}
  \langle x^{2n}(t)\rangle \simeq \frac{(2n){!}}{n!} \left(\frac{D_v}{\gamma^2\mu_1}\right)^n \,t^n.
\end{equation}

Another way to derive the moments of $x(t)$ (\ref{model1_x2n}) is based on the Gaussian distribution of the original process of $x(s)$ in operation time. For a Gaussian process, its PDF can be completely determined from the knowledge of its variance and mean.
Based on the correlation function of $v$ in operation time (\ref{vs_2point}), we calculate the second moment of $x(s)$ in operation time for model (\ref{model1}):
\begin{eqnarray*}
    \langle x^2(s)\rangle &= \int_0^{s}\!\!\!\!\int_0^{s}{\rm d}s_1{\rm d}s_2~ \langle v(s_1)v(s_2)\rangle
    \simeq \frac{2D_v}{\gamma^2}\,s.
\end{eqnarray*}
Since the mean of $v(s)$ is zero, the motion is unbiased and the odd-order moments of $x(s)$ are zero; the even-order moments are
\begin{equation}\label{model1_x2ns}
    \langle x^{2n}(s)\rangle \simeq \frac{(2n){!}}{n!} \left(\frac{D_v}{\gamma^2}\right)^n \,s^n.
\end{equation}
Then using the relation
\begin{equation}\label{model1_x2nts}
  \langle x^{2n}(t)\rangle = \int_0^\infty {\rm d}s~ \langle x^{2n}(s)\rangle h(s,t),
\end{equation}
and the asymptotic expression
\begin{equation*}
  \Phi(\lambda)\simeq \mu_1 \lambda, \quad \rm{as}~ \lambda\rightarrow0,
\end{equation*}
one can also get the result (\ref{model1_x2n}). In the long time, the Langevin system (\ref{model1}) coupled with $\alpha$-dependent subordinator ($1<\alpha<2$) still exhibits normal diffusion (\ref{model1_x2n}) as in the operation time (\ref{model1_x2ns}), although this subordinator might change the PDF of $x(t)$ and $v(t)$ in the Langevin system.

At first glance, the $\alpha$-dependent subordinator ($1<\alpha<2$) has finite mean, and might make no difference with the exponential distribution or simply without any subordinator. This recognition is correct just in some special cases, e.g., the coupled Langevin system (\ref{model1}). But for most of complex system or various statistical quantities, this subordinator may still bring in some new interesting phenomena or diffusion behavior, which essentially depend on whether the observed statistical quantities are related to the multiple-point distribution of the inverse subordinator. For the simple cases, some quantities of the subordinated processes might only depend on the single-point distribution of inverse subordinator $\hat{h}(s,\lambda)$ in (\ref{h_1point}), where
\begin{equation*}
  \Phi(\lambda)\simeq \mu_1 \lambda, \quad \rm{as}~ \lambda\rightarrow0,
\end{equation*}
for long times, just like the procedure (\ref{model1_x2nts}). But if it depends on the two-point distribution of inverse subordinator $\hat{h}(s_2,\lambda_2;s_1,\lambda_1)$ in (\ref{h_2point}), where
\begin{equation*}
  \Phi(\lambda_1)+\Phi(\lambda_2)-\Phi(\lambda_1+\lambda_2) \simeq \mu_\alpha [(\lambda_1+\lambda_2)^\alpha-\lambda_1^\alpha-\lambda_2^\alpha],
\end{equation*}
the result will be different.

In this sense, it is not hard to understand why L\'{e}vy walk exhibits a special sub-ballistic superdiffusion regime when the exponent of waiting times is $1<\alpha<2$ while a trivial phenomenon is observed for CTRWs because of the boundedness of the first moment of the waiting time distribution.  
%like L\'{e}vy flight with the same waiting times distribution. 
The second moment $\langle x^2(t)\rangle$ in L\'{e}vy walk depends on the correlation function of velocity $\langle v(t_1)v(t_2)\rangle$ and thus the two-point distribution $\hat{h}(s_2,\lambda_2;s_1,\lambda_1)$. But $\langle x^2(t)\rangle$ in the case of CTRWs only depends on $\langle x^2(s)\rangle$ and thus depends on single-time distribution $\hat{h}(s,\lambda)$.
So if we consider the overdamped Langevin equation (\ref{model2}) with such a subordinator, the diffusion behaviours for long times will be the same as the ones of the original process.

\section{Numerical simulations of subordinator}\label{Sec6}
Below, we show how to numerically approximate the sample paths of the process $x(t)$ of (\ref{model}). In the first step, we numerically approximate the $\alpha$-dependent subordinator $t(s)$ with $1<\alpha<2$ on the lattice $\{\tau_k=k\Delta \tau:k=1,\cdots,N\}$ where $\Delta \tau=T/N$. For making some preparations, let us give a brief introduction of the idea of two time scales in \cite{BeckerMeerschaertScheffler:2004}.

Suppose that $X_1,X_2,X_3,\cdots$ are the sequence of independent identically distributed positive random variables representing the waiting times between consecutive jumps of the walker, with the distribution \cite{ZaburdaevDenisovKlafter:2015}
\begin{equation}\label{PDF1}
  \phi(t)=\frac{1}{\tau_0}\cdot \frac{\alpha}{(1+t/\tau_0)^{1+\alpha}},  \quad  1<\alpha<2,
\end{equation}
which is consistent to the L\'{e}vy measure $\nu(dy)$ defined in (\ref{nu12}).
The Laplace transform of the PDF $\phi(t)$ is
\begin{equation*}
  \hat{\phi}(\lambda)\simeq  1-\mu_1\lambda  +\mu_\alpha \lambda^\alpha
\end{equation*}
as $\lambda\rightarrow0$.
Note that $X_i$ has a positive mean $\mu_1$.
Consider the total time
\begin{equation*}
  T_{[ct]}=\sum_{i=1}^{[ct]}X_i=\sum_{i=1}^{[ct]}(X_i-\mu_1)+\sum_{i=1}^{[ct]}\mu_1
\end{equation*}
with the scale factor $c$, and $[ct]$ denotes an integer number satisfying $[ct]\leq ct<[ct]+1$.
Note that the first sum grows like $c^{1/\alpha}$ while the second grows like $c$ as $c\rightarrow \infty$. Hence, we cannot get a convergence by normalizing only at one scale. So the L\'{e}vy (and Gaussian) central limit theorem \cite{MeerschaertSikorskii:2011} is not valid here. Instead, we use the technique in \cite{BeckerMeerschaertScheffler:2004} of normalizing $T_{[ct]}$ at two scales, and get the centered and normalized sum
\begin{equation}\label{twoscales}
  T^{c}(t)=c^{-1/\alpha}\sum_{i=1}^{[ct]}(X_i-\mu_1)+c^{-1}\sum_{i=1}^{[ct]}\mu_1.
\end{equation}
Note that $T^{c}(t)$ cannot represent the time of $[ct]$-th jump for large $c$, since it is not non-decreasing. This can be verified from the increment of $T^{c}(t)$ that
for $1<\alpha<2$,
\begin{equation*}
  c^{-1/\alpha}(X_i-\mu_1)+c^{-1}\mu_1\geq0, \quad \rm{only~when}~~ 0<c<1.
\end{equation*}

Taking $c\rightarrow\infty$ in $T^{c}(t)$, we obtain the L\'{e}vy process $T(t)$, but it is not non-decreasing.
Therefore, we consider the non-decreasing supremum process $\bar{T}(t)$ defined as \cite{BeckerMeerschaertScheffler:2004}
\begin{equation}
  \bar{T}(t):= \sup\{T(t'):0<t'<t \}.
\end{equation}
Since the first-passage times (i.e., inverse subordinator defined in (\ref{inverse_st})) of the process $T(t)$ and its supremum process $\bar{T}(t)$ are the same, we will generate the inverse subordinator $s(t)$ based on $T(t)$ in numerical simulations. This is appropriate and can be verified from the double Laplace transform of the PDF of $\bar{T}(t)$ given in \cite{BeckerMeerschaertScheffler:2004,BaeumeraMeerschaert:2007} that
\begin{eqnarray}\label{citeMMM}
    \int_0^\infty\!\!\!\!\int_0^\infty {\rm d}t & {\rm d}_T \mathbb{P}\{\bar{T}(t)<T\}{\rm e}^{-ut}{\rm e}^{-\lambda T}  \nonumber\\
    &= \frac{1-\lambda/q(u)}{u+\mu_1\lambda-\mu_\alpha\lambda^\alpha}
    \simeq \frac{1}{u+\mu_1\lambda-\mu_\alpha\lambda^\alpha},
\end{eqnarray}
as $\lambda\rightarrow0$ and $q$ is a holomorphic function.
Taking the inverse Laplace transform $u\rightarrow t$ of (\ref{citeMMM}), the characteristic function of $\bar{T}(t)$ is obtained as  ${\rm e}^{-t(\mu_1\lambda-\mu_\alpha\lambda^\alpha)}$, consistent to the Laplace exponent $\Phi(\lambda)$ in (\ref{Phi}).

Following the discussions above, we can generate the random variable $X_i$ drawn from the distribution (\ref{PDF1}) by
\begin{equation*}
  X_i=\tau_0[(1-U)^{-1/\alpha}-1],
\end{equation*}
where $U$ is uniformly distributed between $0$ and $1$. Then we get the mean $\mu_1$ of $X_i$:
\begin{equation*}
  \mu_1=\frac{1}{N}\sum_{i=1}^N X_i,
\end{equation*}
and thus obtain the centred and normalized sum as (\ref{twoscales})
\begin{equation}\label{t_lattice}
  t(\tau_k)= \Delta \tau^{1/\alpha}\sum_{i=1}^{k}(X_i-\mu_1)+\Delta \tau\sum_{i=1}^{k}\mu_1,
\end{equation}
which is the approximation value of the subordinator $t(s)$ at lattice $\tau_k, k=1,\cdots,N$.

Based on the subordinator of (\ref{t_lattice}), we use the methods in \cite{Magdziarz:2007} to generate the inverse subordinator process $s(t)$ in (\ref{inverse_st}) and the subordinated process $v(t)=v(s(t))$ in (\ref{model}). Supposing that we have got the velocity $v(t_i)$ on another set of lattices $\{t_i=i\Delta t:i=0,1,\cdots,M\}$, the position $x(t)$ can be obtained directly by
\begin{equation}
  x(t_{i+1})=x(t_i)+v(t_i)\Delta t.
\end{equation}

\begin{figure}
  \centering
  \includegraphics[scale=0.6]{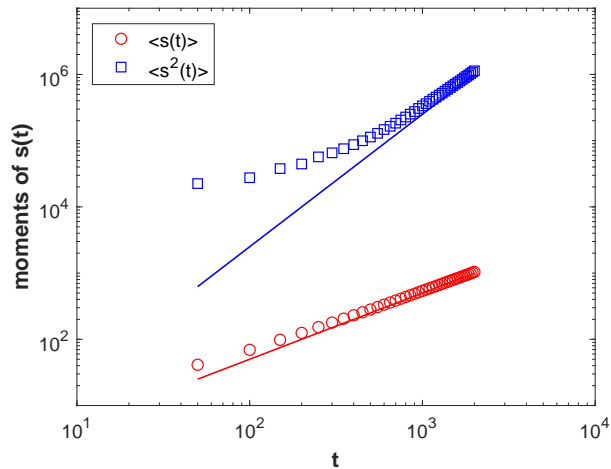}\\
  \caption{First and second moments of inverse subordinator $s(t)$. 1000 trajectories are used with parameters: $T=2000$, $\beta=1.5$,  and $\tau_0=1$. The solid lines denote the theoretical results for long times while the markers are obtained from numerical simulations.}\label{fig3}
\end{figure}

For the inverse subordinator $s(t)$, we simulate its first two moments. From (\ref{h_1point}) and $\Phi(\lambda)\simeq \mu_1\lambda$ for small $\lambda$, we get
\begin{equation}
  h(s,\lambda)\simeq \mu_1 e^{-\mu_1\lambda s},
\end{equation}
and thus
\begin{equation}
  \langle  s(\lambda)\rangle \simeq \frac{1}{\mu_1\lambda^2},\qquad
  \langle  s^2(\lambda)\rangle \simeq \frac{2}{\mu_1^2\lambda^3}.
\end{equation}
Taking inverse Laplace transform $(\lambda\rightarrow t)$ gives
\begin{equation}\label{moments_st}
  \langle  s(t)\rangle \simeq \frac{t}{\mu_1},\qquad
  \langle  s^2(t)\rangle \simeq \frac{t^2}{\mu_1^2}.
\end{equation}
Figure \ref{fig3} shows the numerical simulations of the first and second moments of $s(t)$.
%The simulation of the first and second moments of $s(t)$ is shown in . 
It can be seen that the simulation results (circle markers and square markers) are consistent to the theoretical results (\ref{moments_st}) (solid lines) for long times, which also verifies the long time asymptotic approximation in (\ref{citeMMM}) for subordinator. 

%. This also corresponds to the long time asymptotic way in (\ref{citeMMM}) for subordinator.

\section{Summary and conclusions}\label{Sec7}

L\'{e}vy walk is an important model for describing random walk with finite velocity, which exhibits anomalous superdiffusion phenomenon. For the standard L\'{e}vy walk, it can be divided into three categories, depending on the value of the power-law exponent $\beta$ of the waiting times distribution: ballistic diffusion for $0<\beta<1$, sub-ballistic superdiffusion for $1<\beta<2$, and normal diffusion for $\beta>2$. Based on the feature of finite velocity, we claim that the weakly damped Langevin system in a confined potential together with a subordinator on velocity $v$ can model the dynamics (almost the same as the ones) of L\'{e}vy walk. Friedrich {\it et al.} derived the fractional Klein-Kramers equation retaining retardation effects by master equation of CTRWs in \cite{FriedrichJenkoBauleEule:2006,FriedrichJenkoBauleEule2:2006}, and presented its corresponding Langevin picture in \cite{EuleFriedrichJenkoKleinhans:2007}, where a weakly damped Langevin system is coupled with one-sided $\alpha$-stable subordinator ($0<\alpha<1$). In \cite{FriedrichJenkoBauleEule:2006}, the second moment $\langle x^2(t)\rangle\propto t^2$ was obtained, which is consistent to the ballistic regime of L\'{e}vy walk.
%In the current paper, the intrinsic relation with L\'{e}vy walk is pointed out by investigating the special case of $\alpha$-dependent subordinator with $1<\alpha<2$. 
Another way to characterise L\'{e}vy walk from overdamped Langevin equation is to assume that jump sizes are some functions of waiting times in \cite{MagdziarzSzczotkaZebrowski:2012}.

In this paper, we define a new $\alpha$-dependent subordinator ($1<\alpha<2$) and provide its simulation method when applied to Langevin systems. The weakly damped coupled Langevin equation with this subordinator is build. The lower-order moments of velocity $v(t)$ and position $x(t)$ in this Langevin system are calculated. Especially for $\gamma\neq0$, the diffusion behaviours of the Langevin system are the same as the ones of 
%this Langevin system exhibits the same diffusion behaviour and ergodicity breaking as 
the L\'{e}vy walk in sub-ballistic superdiffusion regime for long times, where strong anomalous diffusion can be observed. There is a relatively intuitive interpretation for this regime in L\'{e}vy walk \cite{ZaburdaevDenisovKlafter:2015}. Compared with the case of $0<\alpha<1$, there are less particles still in their very first flights to form the ballistic fronts when $1<\alpha<2$. So they are slower than the ones of $0<\alpha<1$, but still faster than normal diffusion. Here we present the interpretation from the perspective of Langevin system. For general simple cases, this subordinator might be trivial, working like a linear transform for long times. This is because that the exponent of the single-point distribution of inverse subordinator in Laplace space reduces to be linear with $\lambda$.
But for the model (\ref{model}) with $\gamma\neq0$, the second moment of position $x(t)$ depends on the two-point distribution of inverse subordinator, where $\lambda^\alpha$ plays an important role, and eventually contributes to the sub-ballistic superdiffusion regime.

The overlooked $\alpha$-dependent subordinator with $1<\alpha<2$  helps to model the motion of L\'{e}vy walk in sub-ballistic superdiffusion regime. Keeping this essential/potential mechanism in mind, it will be helpful to characterize more complex stochastic processes with this subordinator, e.g., turbulent in fluids, complex liquids and various biological system. Besides, more complex Langevin system with this subordinator can be considered, such as, the system with a nonlinear external force field, and even the functional distribution of the particle trajectory in the weakly damped or overdamped system.

\section*{Acknowledgments}
This work was supported by the National Natural Science Foundation of China under grant no. 11671182, and the Fundamental Research Funds for the Central Universities under grants no. lzujbky-2018-ot03 and no. lzujbky-2017-ot10.

\appendix
\section{Derivation of (\ref{v4t})}\label{App1}
For simplicity, we denote $h_4(s,t)$ and $g_4(t,s)$ as the four-point distribution of inverse subordinator $s(t)$ and subordinator $t(s)$,  respectively. So the relation of correlation function between operation time $s$ and physical time $t$ is
\begin{equation}\label{A0}
 \fl \langle v(t_4)v(t_3)v(t_2)v(t_1)\rangle =
  \int_0^{\infty}\!\!\!\!\int_0^{{\infty}}\!\!\!\!\int_0^{{\infty}}\!\!\!\!\int_0^{{\infty}} ds_4ds_3ds_2ds_1 ~
  \langle v(s_4)v(s_3)v(s_2)v(s_1)\rangle h_4(s,t).
\end{equation}
On the other hand, the relation of PDF between inverse subordinator $s(t)$ and subordinator $t(s)$ is, similarly to (\ref{h_2point}),
\begin{equation*}
  \hat{h}_4(s,\lambda)=\frac{\partial}{\partial s_1}\frac{\partial}{\partial s_2}\frac{\partial}{\partial s_3}\frac{\partial}{\partial s_4}
  \frac{1}{\lambda_1\lambda_2\lambda_3\lambda_4} \hat{g}_4(\lambda,s).
\end{equation*}
Since (\ref{v4s}) is obtained on the preconditional hypothesis $s_1<s_2<s_3<s_4$, the exact result of (\ref{v4s}) should be written as
\begin{equation}\label{A1}
 \fl \langle v(s_4)v(s_3)v(s_2)v(s_1)\rangle = 4D_v^2\Big(s_1s_3\Theta(s_2-s_1)\Theta(s_4-s_3)+2s_1s_2\Theta(s_3\wedge s_4-s_1\vee s_2)\Big),
\end{equation}
where $\wedge$ denotes minimum and $\vee$ maximum.
Now we claim that the term $s_1s_2\Theta(s_3\wedge s_4-s_1\vee s_2)$ substituted into (\ref{A0}) yields $\langle s(t_1)s(t_2)\rangle\,\Theta(t_3\wedge t_4-t_1\vee t_2)$,
which is sufficient to derive (\ref{v4t}) from (\ref{v4s}).
For convenience, denote the term as $Q(t_4,t_3,t_2,t_1)$. It can be divided into two parts ($s_2>s_1$ and $s_1>s_2$) and written in Laplace space ($t\rightarrow\lambda$) as
\begin{eqnarray}\label{Q41}
 \fl   \hat{Q}(\lambda_4,\lambda_3,\lambda_2,\lambda_1) = \hat{Q}_1(\lambda_4,\lambda_3,\lambda_2,\lambda_1) +\hat{Q}_2(\lambda_4,\lambda_3,\lambda_2,\lambda_1) \nonumber\\
 \fl   = \frac{1}{\lambda_1\lambda_2\lambda_3\lambda_4} \int_0^{\infty}\!\!\!\!\int_0^{{\infty}}\!\!\!\!\int_0^{{\infty}}\!\!\!\!\int_0^{{\infty}}ds_1ds_2ds_3ds_4 ~ s_1s_2 \Theta(s_2-s_1)\Theta(s_3-s_2)\Theta(s_4-s_2)
 \nonumber\\
 \fl ~~  \cdot \frac{\partial}{\partial s_1}\frac{\partial}{\partial s_2}\frac{\partial}{\partial s_3}\frac{\partial}{\partial s_4} \hat{g}_4(\lambda,s)   +\frac{1}{\lambda_1\lambda_2\lambda_3\lambda_4} \int_0^{\infty}\!\!\!\!\int_0^{{\infty}}\!\!\!\!\int_0^{{\infty}}\!\!\!\!\int_0^{{\infty}}ds_1ds_2ds_3ds_4 ~
 \nonumber\\
 \fl   
 ~~ \cdot s_1s_2 \Theta(s_1-s_2)\Theta(s_3-s_1)\Theta(s_4-s_1)
    \frac{\partial}{\partial s_1}\frac{\partial}{\partial s_2}\frac{\partial}{\partial s_3}\frac{\partial}{\partial s_4} \hat{g}_4(\lambda,s).\nonumber
\end{eqnarray}
Through integration by part respect to $s_4$ and $s_3$, the first term $\hat{Q}_1(\lambda_4,\lambda_3,\lambda_2,\lambda_1)$ reduces to
\begin{eqnarray}\label{AQ1}
  \fl  \hat{Q}_1(\lambda_4,\lambda_3,\lambda_2,\lambda_1)  \nonumber\\
 \fl =\frac{1}{\lambda_1\lambda_2\lambda_3\lambda_4} \int_0^{\infty}\!\!\!\!\int_0^{{\infty}}\!\!\!\!\int_0^{{\infty}}\!\!\!\!\int_0^{{\infty}}ds_1ds_2ds_3ds_4 ~  s_1s_2 \Theta(s_2-s_1)\delta(s_3-s_2)\delta(s_4-s_2)
    \frac{\partial}{\partial s_1}\frac{\partial}{\partial s_2} \hat{g}_4(\lambda,s) \nonumber\\
 \fl   =\frac{1}{\lambda_1\lambda_2\lambda_3\lambda_4} \int_0^{\infty}\!\!\!\!\int_0^{{\infty}}ds_1ds_2~  s_1s_2 \Theta(s_2-s_1)
    \frac{\partial}{\partial s_1}\frac{\partial}{\partial s_2} \hat{g}(\lambda_2+\lambda_3+\lambda_4,s_2;\lambda_1,s_1),
\end{eqnarray}
where we have used the formula in the second step that
\begin{eqnarray*}
    \int_0^{\infty}&\!\!\!\!\int_0^{{\infty}}ds_3ds_4\delta(s_3-s_2)\delta(s_4-s_2)\hat{g}_4(\lambda,s)  \\
    &=\int_0^{\infty}\!\!\!\!\int_0^{{\infty}}ds_3ds_4\delta(s_3-s_2)\delta(s_4-s_2) \langle e^{-\lambda_4t(s_4)-\lambda_3t(s_3)-\lambda_2t(s_2)-\lambda_1t(s_1)}\rangle  \\[3pt]
    &=\langle e^{-(\lambda_4+\lambda_3+\lambda_2)t(s_2)-\lambda_1t(s_1)}\rangle  \\[3pt]
    &=\hat{g}(\lambda_2+\lambda_3+\lambda_4,s_2;\lambda_1,s_1).
\end{eqnarray*}
Taking inverse Laplace transform with respect to $\lambda_2,\lambda_3,\lambda_4$ in order gives
\begin{eqnarray*}
  \fl  \frac{1}{\lambda_2\lambda_3\lambda_4}\hat{g}(\lambda_2+\lambda_3+\lambda_4,s_2;\lambda_1,s_1) \overset{\mathcal{L}^{-1}}{\longrightarrow}
    \int_0^{t_2}\!\!\!\!\int_0^{t_3}\!\!\!\!\int_0^{t_4}dt_2'dt_3'dt_4' ~  \delta(t_2'-t_3')\delta(t_2'-t_4') \hat{g}(t_2',s_2;\lambda_1,s_1) \\
    = \int_0^{t_2}dt_2' ~  \hat{g}(t_2',s_2;\lambda_1,s_1)
    \overset{\mathcal{L}_{t_2\rightarrow\lambda_2}}{\longrightarrow}  \frac{1}{\lambda_2} \hat{g}(\lambda_2,s_2;\lambda_1,s_1),
\end{eqnarray*}
when $t_3\wedge t_4>t_2$.
Substituting it into (\ref{AQ1}), we have
\begin{eqnarray*}
  \hat{Q}_1(t_4,t_3,\lambda_2,\lambda_1)&=\frac{1}{\lambda_1\lambda_2} \int_0^{\infty}\!\!\!\!\int_0^{{\infty}}ds_1ds_2~  s_1s_2 \Theta(s_2-s_1)
    \frac{\partial}{\partial s_1}\frac{\partial}{\partial s_2} \hat{g}(\lambda_2,s_2;\lambda_1,s_1) \\
    &= \int_0^{\infty}\!\!\!\!\int_0^{{\infty}}ds_1ds_2 ~ s_1s_2 \Theta(s_2-s_1) \hat{h}(s_2,\lambda_2;s_1,\lambda_1).
\end{eqnarray*}
Similarly, $\hat{Q}_2(t_4,t_3,\lambda_2,\lambda_1)$ can be obtained when $t_3\wedge t_4>t_1$. Therefore,
\begin{equation*}
  \hat{Q}(t_4,t_3,\lambda_2,\lambda_1)=\int_0^{\infty}\!\!\!\!\int_0^{{\infty}}ds_1ds_2 ~ s_1s_2 \hat{h}(s_2,\lambda_2;s_1,\lambda_1),
\end{equation*}
when $t_3\wedge t_4>t_1\vee t_2$.
Taking inverse Laplace transform with respect to $\lambda_2,\lambda_1$ gives
\begin{equation*}
  Q(t_4,t_3,t_2,t_1)= \langle s(t_1)s(t_2)\rangle \, \Theta(t_3\wedge t_4-t_1\vee t_2).
\end{equation*}

\section{Correlation functions of $v(t)$ and $x(t)$ from (\ref{v2lam22}) and (\ref{x2lam2})}\label{App2}

Here we derive the correlation functions of $v(t)$ and $x(t)$ by inversing (\ref{v2lam22}) and (\ref{x2lam2}), respectively.
Since $\lambda_1,\lambda_2\rightarrow0$, we only give the results for long times
and further assume $t_1<t_2$ without loss of generality. Taking the inverse Laplace transform ($\lambda_1\rightarrow t_1,\lambda_2\rightarrow t_2$) of the three terms in (\ref{v2lam22}),  respectively, yields
\begin{eqnarray*}
    \mathcal{L}^{-1}\left[\frac{(\lambda_1+\lambda_2)^{\alpha}}{\lambda_1\lambda_2(\lambda_1+\lambda_2)}\right]
    = \mathcal{L}^{-1} \left[ \frac{(\lambda_1+\lambda_2)^{\alpha-2}}{\lambda_1}+\frac{(\lambda_1+\lambda_2)^{\alpha-2}}{\lambda_2}\right]
    = \frac{t_1^{1-\alpha}}{\Gamma(2-\alpha)} ,
\end{eqnarray*}
\begin{eqnarray*}
    \mathcal{L}^{-1}\left[\frac{\lambda_1^{\alpha}}{\lambda_1\lambda_2(\lambda_1+\lambda_2)}\right]
    =\mathcal{L}^{-1}\left[\frac{\lambda_1^{\alpha-2}}{\lambda_2}-\frac{\lambda_1^{\alpha-2}}{\lambda_1+\lambda_2}\right]
    =\frac{t_1^{1-\alpha}}{\Gamma(2-\alpha)},
\end{eqnarray*}
\begin{eqnarray*}
    \mathcal{L}^{-1}\left[\frac{\lambda_2^{\alpha}}{\lambda_1\lambda_2(\lambda_1+\lambda_2)}\right]
    =\mathcal{L}^{-1}\left[\frac{\lambda_2^{\alpha-2}}{\lambda_1}-\frac{\lambda_2^{\alpha-2}}{\lambda_1+\lambda_2}\right]
    =\frac{t_2^{1-\alpha}}{\Gamma(2-\alpha)}-\frac{(t_2-t_1)^{1-\alpha}}{\Gamma(2-\alpha)}.
\end{eqnarray*}
Note that $t_1<t_2$ has been used in the inverse Laplace transform of these three terms. Then the correlation function of $v(t)$ is obtained as
\begin{equation}\label{correlation_v}
  \langle v(t_1)v(t_2)\rangle= \frac{D_v\mu_\alpha}{\gamma\mu_1}\cdot \frac{(t_2-t_1)^{1-\alpha}-t_2^{1-\alpha}}{\Gamma(2-\alpha)}.
\end{equation}

Similarly, the correlation function of $x(t)$ is
\begin{eqnarray*}
  \langle x(t_1)x(t_2)\rangle \simeq \frac{D_v\mu_\alpha}{\gamma\mu_1} \cdot \left[
  \frac{\alpha}{\Gamma(4-\alpha)}t_1^{3-\alpha}-\frac{t_1}{\Gamma(3-\alpha)}\Big(t_2^{2-\alpha}-(t_2-t_1)^{2-\alpha}\Big)\right. \\
    ~~~~~~~~~~~~~~~~~~~~\left.+\frac{t_1^2t_2^{1-\alpha}}{2\Gamma(2-\alpha)}\,{}_2F_1(\alpha-1,2;3;t_1/t_2)\right],
\end{eqnarray*}
where we have used the formula \cite{GradshteynRyzhikGeraniumsTseytlin:1980}
\begin{equation*}
  \int_0^u\frac{x^{\mu-1}}{(1+\beta x)^\nu}dx = \frac{u^\mu}{\mu}\,{}_2F_1(\nu,\mu;1+\mu;-\beta u),  \quad |\rm{arg}(1+\beta u)|<\pi, \Re(\mu)>0,
\end{equation*}
in the calculation of inverse Laplace transform.
Fixing $t_1$ and letting $t_2\rightarrow\infty$, we get
\begin{equation}
  \langle x(t_1)x(t_2)\rangle \simeq \frac{D_v\mu_\alpha}{\gamma\mu_1} \frac{\alpha}{\Gamma(4-\alpha)}t_1^{3-\alpha}.
\end{equation}
For $t_1=t_2=t$, we have
\begin{equation*}
  _2F_1(\alpha-1,2;3;1)=\frac{2\Gamma(2-\alpha)}{\Gamma(4-\alpha)},
\end{equation*}
and thus
\begin{equation}\label{Bx2}
  \langle x^2(t)\rangle \simeq \frac{D_v\mu_\alpha}{\gamma\mu_1}\frac{2\alpha-2}{\Gamma(4-\alpha)}\,t^{3-\alpha}.
\end{equation}

\section*{References}
\bibliographystyle{iopart-num}
\bibliography{Reference}

\end{document}